\documentclass[12pt]{article}
\usepackage[reqno]{amsmath}
\usepackage{bbm}
\usepackage{epsfig}
\usepackage{array}
\usepackage{float}

\usepackage{a4}

\usepackage{a4wide}
\usepackage{wasysym}

\def\be{\begin{equation}}
\def\ee{\end{equation}}
\def\gs{\mathrel{
   \rlap{\raise 0.511ex \hbox{$>$}}{\lower 0.511ex \hbox{$\sim$}}}}
\def\ls{\mathrel{
   \rlap{\raise 0.511ex \hbox{$<$}}{\lower 0.511ex \hbox{$\sim$}}}}

\newcommand{\ba}{\begin{array}{c}}
\newcommand{\baz}{\begin{array}{cc}}
\newcommand{\bad}{\begin{array}{ccc}}
\newcommand{\bea}{\begin{equation} \begin{array}{c}}
\newcommand{\eea}{ \end{array} \end{equation}}
\newcommand{\ea}{\end{array}}

\newcommand{\meff}{\mbox{$\langle m \rangle$}}

\begin{document}

\title{
\vspace{-2cm}
\hfill {\framebox{\small\bf hep-ph/0408195}}\\
\vspace{-0.3cm}
\hfill {\small\bf SISSA 57/2004/EP}\\
\vspace{-0.3cm}
\hfill {\small\bf BIHEP-TH-2004-17}\\
\vskip 0.8cm
\bf Flavor Democracy and Type-II Seesaw Realization of Bilarge Neutrino Mixing 
}
\vspace{0.5cm}
\author{
{\large\bf Werner Rodejohann}\thanks{email: \tt werner@sissa.it} $^a$
$\;$ and $\;$ 
{\large\bf Zhi-zhong Xing}\thanks{email: \tt xingzz@mail.ihep.ac.cn} $^b$
\\ \\ 
{\normalsize \it $^a$ Scuola Internazionale Superiore di Studi Avanzati}
{\normalsize \it Via Beirut 2--4,} \\
{\normalsize \it I-34014 Trieste, Italy}\\
{\normalsize \it and}
{\normalsize \it Istituto Nazionale di Fisica Nucleare}
{\normalsize \it Sezione di Trieste,} \\
{\normalsize \it  I-34014 Trieste, Italy}\\ \\ 
{\normalsize \it $^b$ CCAST (World Laboratory), P.O. Box 8730,
Beijing 100080, China} \\
{\normalsize \it ~ and Institute of High Energy Physics, 
Chinese Academy of Sciences,}\\ 
{\normalsize \it P.O. Box 918 (4), Beijing 100039, China}
}
\date{}
\maketitle
\thispagestyle{empty}
\vspace{0.4cm}
\begin{abstract}
\noindent 
We generalize the democratic neutrino mixing Ansatz by incorporating
the type-II seesaw mechanism with S(3) flavor symmetry. 
For only the triplet mass term or only the conventional seesaw term large 
neutrino mixing can be achieved only by assuming an 
unnatural suppression of the flavor democracy contribution. 
We show that bilarge neutrino mixing can naturally 
appear if the flavor democracy term is strongly 
suppressed due to significant cancellation 
between the conventional seesaw and triplet mass terms.   
Explicit S(3) symmetry breaking yields successful neutrino phenomenology
and various testable correlations between the neutrino mass and mixing 
parameters. Among the results are a normal neutrino mass ordering, 
$0.005 \le |U_{e3}| \le 0.057$, $1 - \sin^2 2\theta_{23} \ge 0.005$, 
positive $J_{\rm CP}$ and moderate cancellation in the effective
mass of the neutrinoless double beta decay.  
\end{abstract}

\newpage


The elegant Super-Kamiokande \cite{SK}, SNO \cite{SNO}, K2K \cite{K2K}
and KamLAND \cite{KM} experiments have provided us with very 
convincing evidence that the long-standing solar neutrino 
deficit and the atmospheric neutrino anomaly are both due to 
neutrino oscillations, which can naturally occur if neutrinos are 
massive and lepton flavors are mixed. A big puzzle is that the mass
scale of three active neutrinos (i.e., $\nu_e$, $\nu_\mu$ and $\nu_\tau$) 
is extremely low, at most of ${\cal O}(0.1)$ eV. In addition, 
lepton flavor mixing involves two remarkably large 
angles, $\theta_{12} \sim 33^\circ$ and
$\theta_{23} \sim 45^\circ$ in the standard parametrization.   
To understand the smallness of neutrino masses, a number of theoretical 
and phenomenological ideas have been proposed in the 
literature \cite{Review1}. Among them, the most natural idea is the 
seesaw mechanism \cite{SS}. While the seesaw mechanism itself can 
qualitatively explain why neutrino masses are so small, it is unable 
to make any concrete predictions unless a specific lepton flavor structure 
is assumed. Hence an appropriate combination of the seesaw mechanism and 
possible flavor symmetries \cite{FS} or texture zeros \cite{TZ} is 
practically needed, in order to quantitatively account for the neutrino
mass spectrum and the bilarge lepton mixing pattern. Some interesting
attempts in this direction \cite{Review2} have been made recently. 

\vspace{0.2cm}

In this letter we aim to interpret current experimental data on neutrino
masses and lepton flavor mixing angles by incorporating the type-II seesaw 
mechanism \cite{typeII} with S(3) flavor 
symmetry and its explicit breaking. Our physical
motivation is rather simple. The charged lepton
mass matrix with $\rm S(3)_L \times S(3)_R$ symmetry
(i.e., flavor democracy) and the effective Majorana neutrino mass matrix 
with S(3) permutation symmetry may in general be written as
\begin{eqnarray}
M^{(0)}_l & = & \frac{c^{~}_l}{3} \left (\begin{matrix}
1       & 1     & 1 \cr
1       & 1     & 1 \cr
1       & 1     & 1 \cr \end{matrix} \right ) \; ,
\nonumber \\
M^{(0)}_\nu & = & c_\nu \left [ \left (\begin{matrix}
1       & 0     & 0 \cr
0       & 1     & 0 \cr
0       & 0     & 1 \cr \end{matrix} \right ) + 
r_\nu \left (\begin{matrix}
1       & 1     & 1 \cr
1       & 1     & 1 \cr
1       & 1     & 1 \cr \end{matrix} \right ) \right ] \; ,
\end{eqnarray}
in which $c^{~}_l$ and $c_\nu$ measure the corresponding mass scales of 
charged leptons and light neutrinos, and $r_\nu$ is in principle an 
arbitrary parameter. A soft breakdown of the above permutation symmetry
can lead to realistic lepton mass matrices
$M_l = M^{(0)}_l + \Delta M_l$ and $M_\nu = M^{(0)}_\nu + \Delta M_\nu$
with proper mass eigenvalues. Then the lepton flavor mixing matrix $U$ 
arises from the mismatch between the diagonalization of $M_l$ and
that of $M_\nu$. It has been noticed in 
Refs.\ \cite{FX96,Tanimoto1,Tanimoto2,Branco,FX04} that $r_\nu$  
must be vanishing or strongly suppressed such that a bilarge
neutrino mixing pattern can be generated. In the spirit of 't Hooft's 
naturalness principle \cite{Hooft}, however,  
$|r_\nu| = {\cal O}(1)$ seems more likely than $r_\nu =0$ or 
$|r_\nu| \ll 1$. The point will become clear when the smallness
of $c_\nu$ is attributed to the seesaw mechanism. We find that the
conventional (type-I) seesaw mechanism cannot help out 
(see also \cite{Branco}), but the 
type-II seesaw scenario may provide a natural interpretation of
small neutrino masses and bilarge lepton mixing angles even in the
case of $|r_\nu| = {\cal O}(0.1)$ to ${\cal O}(1)$. 

\vspace{0.2cm}

In type-II seesaw models with three right-handed neutrinos, 
the neutrino mass term reads 
\begin{equation}
- {\cal L}_{\rm mass} \; = \; \frac{1}{2} 
\overline{(\nu, ~\nu^{\rm c} )^{~}_{\rm L}}
\left ( \begin{matrix}
M_{\rm L}     & M_{\rm D} \cr
M^T_{\rm D}    & M_{\rm R} \cr \end{matrix} \right ) 
\left ( \begin{matrix}
\nu^{\rm c} \cr \nu \cr \end{matrix} \right )_{\rm R} \; ,
\end{equation}
where $\nu$ denotes the column vector of three neutrino fields,
$M_{\rm D}$ stands for the $3\times 3$ Dirac neutrino mass matrix,
$M_{\rm L}$ and $M_{\rm R}$ represent the symmetric $3\times 3$
mass matrices of left-handed and right-handed Majorana 
neutrinos respectively. As $M_{\rm L}$ results from a 
$\rm SU(2)_L$ triplet term of the Yukawa interactions.
its scale might be considerably lower 
than the gauge symmetry breaking scale $v \approx 174$ GeV. On
the other hand, the scale of $M_{\rm R}$ can naturally be much
higher than $v$, because right-handed neutrinos are $\rm SU(2)_L$
singlets and their corresponding mass term is not subject to 
gauge symmetry breaking. The strong hierarchy between the 
scales of $M_{\rm R}$ and $M_{\rm L}$ or $M_{\rm D}$ allow us to make 
some safe approximations in diagonalizing the $6\times 6$ neutrino mass 
matrix in Eq.\ (2) and arrive at an effective mass matrix for three
light (essentially left-handed) neutrinos \cite{typeII}:
\begin{equation}
M_\nu \approx M_{\rm L}  -  M_{\rm D} M^{-1}_{\rm R} M^T_{\rm D} \; .
\end{equation}
For a phenomenological study 
of neutrino masses and lepton flavor mixing, we 
assume a discrete left-right symmetry between $M_{\rm L}$ and 
$M_{\rm R}$, whose mass scales are characterized respectively by the 
vacuum expectation values (vevs) of two triplet fields, $v^{~}_{\rm L}$ and
$v^{~}_{\rm R}$. Consequently, the usual left-right symmetric relation 
$v^{~}_{\rm L} v^{~}_{\rm R} = \gamma v^2$ holds, 
where $\gamma$ is a model-dependent factor of 
${\cal O}(1)$. 
As investigated recently, the interplay of the two terms in the 
type-II seesaw formula can result in several interesting effects. 
One can, e.g., upgrade a hierarchical neutrino 
mass spectrum to a quasi-degenerate one \cite{anki} or create 
deviations from the bimaximal neutrino mixing pattern \cite{WR04}.  
In this letter we shall take advantage of possible cancellations 
hidden in the type-II seesaw mechanism, which is an intriguing feature 
when the two mass terms on the right-hand 
side of Eq.\ (3) contribute to $M_\nu$ with comparable magnitudes. 

\vspace{0.2cm}

Imposing S(3) flavor symmetry on $M_{\rm L}$ and $M_{\rm R}$ and 
allowing for soft symmetry breaking, we write down
\begin{eqnarray}
M_{\rm L} & = & v^{~}_{\rm L} \left [ \left (\begin{matrix}
1       & 0     & 0 \cr
0       & 1     & 0 \cr
0       & 0     & 1 \cr \end{matrix} \right ) + 
r_\nu \left (\begin{matrix}
1       & 1     & 1 \cr
1       & 1     & 1 \cr
1       & 1     & 1 \cr \end{matrix} \right ) \right ] + 
\Delta M_{\rm L} \; ,
\nonumber \\
M_{\rm R} & = & v^{~}_{\rm R} \left [ \left (\begin{matrix}
1       & 0     & 0 \cr
0       & 1     & 0 \cr
0       & 0     & 1 \cr \end{matrix} \right ) + 
r_\nu \left (\begin{matrix}
1       & 1     & 1 \cr
1       & 1     & 1 \cr
1       & 1     & 1 \cr \end{matrix} \right ) \right ] +
\Delta M_{\rm R} \; .
\end{eqnarray}
On the other hand, the flavor democracy or $\rm S(3)_L \times S(3)_R$ 
symmetry can be imposed on the Dirac neutrino mass matrix $M_{\rm D}$ 
and the charged lepton mass matrix $M_l$, whose eigenvalues appear
to be hierarchical as those of up- or down-type quarks \cite{Koide}. 
Once soft symmetry breaking is 
taken into account, $M_{\rm D}$ and $M_l$ read 
\begin{eqnarray}
M_{\rm D} & = & \frac{c^{~}_{\rm D}}{3} \left (\begin{matrix}
1       & 1     & 1 \cr
1       & 1     & 1 \cr
1       & 1     & 1 \cr \end{matrix} \right ) + \Delta M_{\rm D} \; ,
\nonumber \\
M_l & = & ~ \frac{c^{~}_l}{3} \left (\begin{matrix}
1       & 1     & 1 \cr
1       & 1     & 1 \cr
1       & 1     & 1 \cr \end{matrix} \right ) + \Delta M_l \;\; . \;\;\;\;\;
\end{eqnarray}   
To make concrete predictions, one has to specify the patterns of
$\Delta M_{\rm L}$, $\Delta M_{\rm R}$, $\Delta M_{\rm D}$ and
$\Delta M_l$. For the sake of simplicity, we follow 
Refs.\ \cite{FX96} and \cite{FX04} to take
\begin{eqnarray}
\Delta M_{\rm L} & = & v^{~}_{\rm L} \left ( \begin{matrix}
-\delta_{\rm M} & 0 & 0 \cr
0 & +\delta_{\rm M} & 0 \cr
0 & 0 & \varepsilon^{~}_{\rm M} \cr \end{matrix} \right ) \; ,
\nonumber \\
\Delta M_{\rm R} & = & v^{~}_{\rm R} \left ( \begin{matrix}
-\delta_{\rm M} & 0 & 0 \cr
0 & +\delta_{\rm M} & 0 \cr
0 & 0 & \varepsilon^{~}_{\rm M} \cr \end{matrix} \right ) \; ,
\end{eqnarray}
where left-right symmetry has been implemented. For the  Dirac fermion 
sector, we choose 
\begin{eqnarray}
\Delta M_{\rm D} & = & \frac{c^{~}_{\rm D}}{3} \left ( \begin{matrix}
-i\delta_{\rm D} & 0 & 0 \cr
0 & +i\delta_{\rm D} & 0 \cr
0 & 0 & \varepsilon^{~}_{\rm D} \cr \end{matrix} \right ) \; ,
\nonumber \\
\Delta M_l & = & ~ \frac{c^{~}_l}{3} \left ( \begin{matrix}
-i\delta_l & 0 & 0 \cr
0 & +i\delta_l ~ & 0 \cr
0 & 0 & ~ \varepsilon^{~}_l \cr \end{matrix} \right ) \; . 
\end{eqnarray}
Note that $\delta_{{\rm M, D},l}$ and $\varepsilon^{~}_{{\rm M, D},l}$
are small perturbative parameters and their magnitudes are at most
of ${\cal O}(0.1)$. Note also that we have introduced imaginary 
perturbations in $\Delta M_{\rm D}$ and $\Delta M_l$, in order to
accommodate leptonic CP violation. Calculating the effective neutrino
mass matrix $M_\nu$ by using Eqs.\ (3)--(7), we obtain
\begin{eqnarray}
M_\nu & \approx & v^{~}_{\rm L} \left [ \left ( \begin{matrix}
1 & 0 & 0 \cr
0 & 1 & 0 \cr
0 & 0 & 1 \cr \end{matrix} \right ) + r_\nu \left ( \begin{matrix}
1 & 1 & 1 \cr
1 & 1 & 1 \cr
1 & 1 & 1 \cr \end{matrix} \right ) + \left ( \begin{matrix}
-\delta_{\rm M} & 0 & 0 \cr
0 & +\delta_{\rm M} & 0 \cr
0 & 0 & \varepsilon^{~}_{\rm M} \cr \end{matrix} \right ) \right ]
\nonumber \\
&  & - \frac{\tilde{c}^2_{\rm D}}{v^{~}_{\rm R}}
\left [ \left ( 1 - \tilde{\varepsilon}^{~}_{\rm M} \right )
\left ( \begin{matrix}
1 & 1 & 1 \cr
1 & 1 & 1 \cr
1 & 1 & 1 \cr \end{matrix} \right ) + \frac{1}{3} \left ( \begin{matrix}
-2i\delta_{\rm D} & 0 & \varepsilon^{~}_{\rm D} - i\delta_{\rm D} \cr
0 & +2i\delta_{\rm D} & \varepsilon^{~}_{\rm D} + i\delta_{\rm D} \cr
\varepsilon^{~}_{\rm D} - i\delta_{\rm D} & 
\varepsilon^{~}_{\rm D} + i\delta_{\rm D} & 2 \varepsilon^{~}_{\rm D} \cr
\end{matrix} \right ) \right ] \; ,
\end{eqnarray}
where $\tilde{c}^{~}_{\rm D} \equiv c^{~}_{\rm D}/\sqrt{3 (1 + 3 r_\nu)}$,
$\tilde{\varepsilon}^{~}_{\rm M} \equiv \varepsilon^{~}_{\rm M}/
[3 (1 + 3r_\nu)]$, and terms of ${\cal O}(\delta^2_{\rm M,D})$ and
${\cal O}(\varepsilon^2_{\rm M,D})$ have been neglected.
It is quite obvious that the matrices
proportional to $v^{~}_{\rm L}$ and $\tilde{c}^2_{\rm D}/v^{~}_{\rm R}$
in Eq.\ (8) arise respectively from $M_{\rm L}$ and 
$M_{\rm D} M^{-1}_{\rm R} M^T_{\rm D}$. Their relative contributions to
$M_\nu$ can be classified into three typical cases:
\begin{itemize}
\item       In the limit of $v^{~}_{\rm L} \rightarrow 0$, we are left
with the conventional (type-I) seesaw result of $M_\nu$, whose leading
term displays flavor democracy. Because both $M_l$ and $M_\nu$ come from
the explicit (soft) breaking of flavor democracy in this special case
(similar to the case of democratic quark mass matrices \cite{Review1}), 
no large lepton flavor mixing can appear. To suppress or avoid such a 
flavor democracy term in the type-I seesaw expression of $M_\nu$ 
(and thereby to open the possibility of generating large 
neutrino mixing angles), other possible flavor symmetries (such as 
$\rm Z_3$ symmetry \cite{Branco}) have to be taken into account. 
\item       In the limit of 
$\tilde{c}^2_{\rm D}/v^{~}_{\rm R} \rightarrow 0$,
we obtain $M_\nu \approx M_{\rm L}$. This pure triplet case can
accommodate current experimental data of solar and atmospheric neutrino
oscillations, if $\varepsilon^{~}_{\rm M} \gg \delta_{\rm M} \sim
r_\nu$ is satisfied \cite{FX04}. To be more specific, 
$r_\nu/\varepsilon^{~}_{\rm M} \sim 6.1 \times 10^{-3}$ has been
obtained in Ref.\ \cite{FX04} without any fine-tuning. 
As $\varepsilon^{~}_{\rm M} = {\cal O}(0.1)$ is most plausible,
the magnitude of $r_\nu$ must be of ${\cal O}(10^{-3})$ or
${\cal O}(10^{-4})$. Such a small result implies that the two S(3) 
symmetry terms in $M_{\rm L}$ are not balanced --- one of them (i.e., 
the flavor democracy term) is strongly suppressed. This seems 
unnatural in some sense, since $|r_\nu| = {\cal O}(1)$ is more or
less expected from the point of view of 't Hooft's naturalness principle.
\item       The two mass terms of $M_\nu$ in Eq.\ (8) are comparable
in magnitude and lead to significant cancellation. A particularly
interesting possibility is that the two flavor democracy terms,
which are proportional to $r_\nu$ and 
$(1- \tilde{\varepsilon}^{~}_{\rm M})$ respectively, 
may essentially cancel each other. In this case, 
\begin{equation}
r_\nu \approx \frac{\tilde{c}^2_{\rm D}}{v^{~}_{\rm L} v^{~}_{\rm R}}
\left (1 - \tilde{\varepsilon}^{~}_{\rm M} \right ) =
\frac{\tilde{c}^2_{\rm D}}{\gamma v^2}
\left (1 - \tilde{\varepsilon}^{~}_{\rm M} \right ) \; 
\end{equation}
is likely to be of ${\cal O}(0.1)$ to ${\cal O}(1)$ (e.g., 
$c^{~}_{\rm D} \sim m^{~}_t \approx v$ might hold in a specific 
GUT framework with lepton-quark symmetry, such as some SO(10) models). 
We carry out a careful 
numerical analysis of this typical type-II seesaw scenario
and find that the bilarge neutrino mixing pattern can actually be 
reproduced without fine-tuning. Before presenting our numerical
results, we would like to give some more comments on the
consequences of Eq.\ (9).
\end{itemize}

Note that the possibility of $r_\nu \sim -1/3$, which may significantly
enhance the magnitude of $\tilde{\varepsilon}^{~}_{\rm M}$, is found
to be disfavored in fitting current neutrino oscillation data. 
In the following we will constrain ourselves to positive 
and small perturbative parameters.  
With the definition $\zeta_\nu \equiv c^2_{\rm D}/(\gamma v^2)$, from which
$\tilde{c}^2_{\rm D}/(\gamma v^2) = \zeta_\nu/[3(1 + 3r_\nu)]$ can be
expressed, we then obtain 
\begin{equation} \label{eq:zetar}
r_\nu \approx \frac{1}{6} \left ( -1 \pm \sqrt{1 + 4 \zeta_\nu}
\right ) \; 
\end{equation}
by solving Eq.\ (9) in the leading-order approximation (i.e., in the
neglect of $\tilde{\varepsilon}^{~}_{\rm M}$).
This rough result clearly shows that $|r_\nu|$ is most likely to be
of ${\cal O}(0.1)$ to ${\cal O}(1)$, provided 
$\zeta_\nu = {\cal O}(1)$ holds. 
Typically, taking for instance $\zeta_\nu = 2$, we arrive at
$r_\nu \approx 1/3$ or $r_\nu \approx -2/3$. Now the question is whether 
in the outlined framework bilarge neutrino mixing can be achieved. 
Inserting Eq.\ (10) into (8) gives 
\begin{equation}
M_\nu \approx v^{~}_{\rm L} \left ( \begin{matrix}
1-\delta_{\rm M} - 2i \hat{\delta}_{\rm D} 
& 0 & -\hat{\varepsilon}^{~}_{\rm D} + i\hat{\delta}_{\rm D} \cr
0 & 1+\delta_{\rm M} + 2i\hat{\delta}_{\rm D} 
& -\hat{\varepsilon}^{~}_{\rm D} -i\hat{\delta}_{\rm D} \cr
-\hat{\varepsilon}^{~}_{\rm D} + i\hat{\delta}_{\rm D} &
-\hat{\varepsilon}^{~}_{\rm D} - i\hat{\delta}_{\rm D} &
1 + \varepsilon^{~}_{\rm M} - 2\hat{\varepsilon}^{~}_{\rm D} 
\cr \end{matrix} \right ) \; ,
\end{equation}
where $\hat{\varepsilon}^{~}_{\rm D} \equiv 
\varepsilon^{~}_{\rm D} \zeta_\nu /[9(1+3r_\nu)]$ and
$\hat{\delta}_{\rm D} \equiv \delta_{\rm D}
\zeta_\nu /[9(1+3r_\nu)]$. One may diagonalize this symmetric
mass matrix by the transformation
$U_\nu M_\nu U^T_\nu = {\rm Diag} \{ m_1, m_2, m_3 \}$, where
$U_\nu$ is a unitary matrix and $m_i$ (for $i=1,2,3$) denote
the physical masses of three light neutrinos. It is obvious
that $m_1 \approx m_2 \approx m_3$ must hold to leading order. 
The observed solar and atmospheric neutrino mass-squared 
differences $\Delta m^2_{\odot} \equiv \Delta m^2_{21} 
\sim 10^{-5} ~{\rm eV}^2$ and
$\Delta m^2_{\rm A} \equiv \Delta m^2_{32} \sim 10^{-3} ~{\rm eV}^2$ 
are proportional to $v^2_{\rm L}$, and their different magnitudes
are governed by the relevant perturbative parameters 
($\delta_{\rm M}$, $\varepsilon^{~}_{\rm M}$, etc). The presence
of $\hat{\varepsilon}^{~}_{\rm D}$ and $\hat{\delta}_{\rm D}$ 
makes it possible to generate suitable rotation angles in $U_\nu$.
The mismatch between $U_\nu$ and the unitary matrix $U_l$, which is
defined to diagonalize $M_l$ (i.e.,
$U_l M_l U^T_l = {\rm Diag}\{ m_e, m_\mu, m_\tau \}$) and given 
by \cite{FX04}
\begin{equation}
U_l \approx \left ( \begin{matrix}
\frac{1}{\sqrt{2}} & \frac{-1}{\sqrt{2}} & 0 \cr\cr
\frac{1}{\sqrt{6}} & \frac{1}{\sqrt{6}} & \frac{-2}{\sqrt{6}} \cr\cr
\frac{1}{\sqrt{3}} & \frac{1}{\sqrt{3}} & \frac{1}{\sqrt{3}} \cr
\end{matrix} \right ) +
i \sqrt{\frac{m_e}{m_\mu}}
\left ( \begin{matrix}
\frac{1}{\sqrt{6}}      & ~ \frac{1}{\sqrt{6}} ~        &
\frac{-2}{\sqrt{6}} \cr\cr
\frac{1}{\sqrt{2}}      & ~ \frac{-1}{\sqrt{2}} ~       & 0 \cr\cr
0       & ~ 0 ~ & 0 \cr \end{matrix} \right )
+ \frac{m_\mu}{m_\tau} \left ( \begin{matrix}
0       & 0     & 0 \cr\cr
\frac{1}{\sqrt{6}}      & \frac{1}{\sqrt{6}}    & \frac{1}{\sqrt{6}} \cr\cr
\frac{-1}{\sqrt{12}}    & \frac{-1}{\sqrt{12}}  & \frac{1}{\sqrt{3}}
\cr \end{matrix} \right ) \; ,
\end{equation}
measures the strength of lepton flavor mixing --- namely, 
$U = U_l U^\dagger_\nu$. Although a bilarge neutrino mixing pattern
is naturally expected from this democratic type-II seesaw scenario, 
we find it very difficult to obtain a simple analytical 
expression of $U_\nu$ to make the result of $U$ more transparent. In
this case, we shall do a numerical analysis of our phenomenological
Ansatz without sticking to the condition given in Eq.\ (9) or (10).

\vspace{0.2cm}

\begin{figure}
\begin{center}
\vspace{-5cm}
\epsfig{file=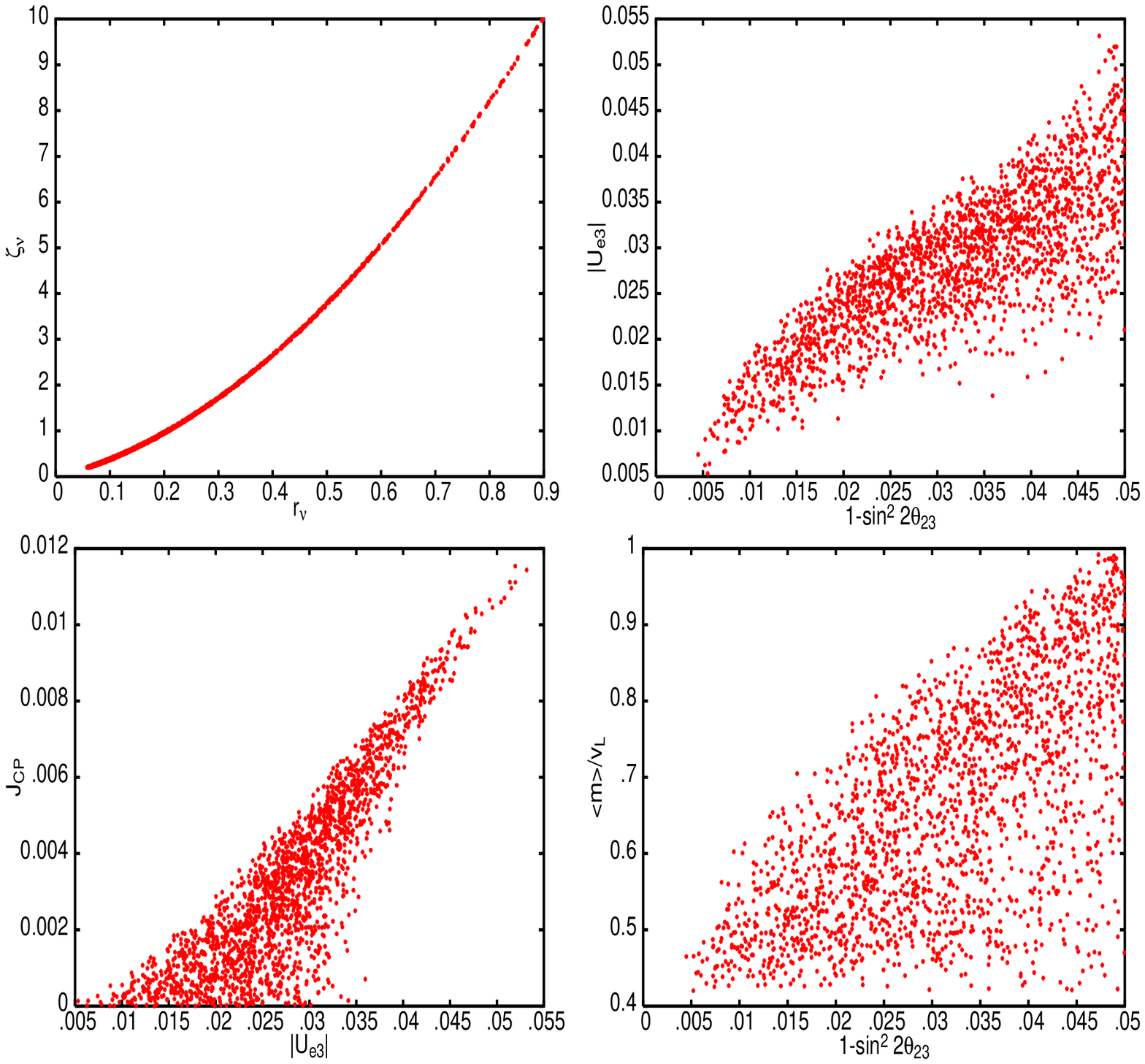,width=16cm,height=24cm}
\vspace{-6.8cm}
\caption{\label{fig:crbel}Scatter plot of the parameters 
$\zeta_\nu$ against $r_\nu$ as well as 
$1 - \sin^2 2 \theta_{23}$ against $|U_{e3}|$, 
$|U_{e3}|$ against $J_{\rm CP}$ and $1 - \sin^2 2 \theta_{23}$ against 
$\langle m \rangle/v^{~}_{\rm L}$ for the case of arbitrary 
$\zeta_\nu$ and $r_\nu$.} 
\end{center}
\end{figure}

\begin{figure}
\begin{center}
\epsfig{file=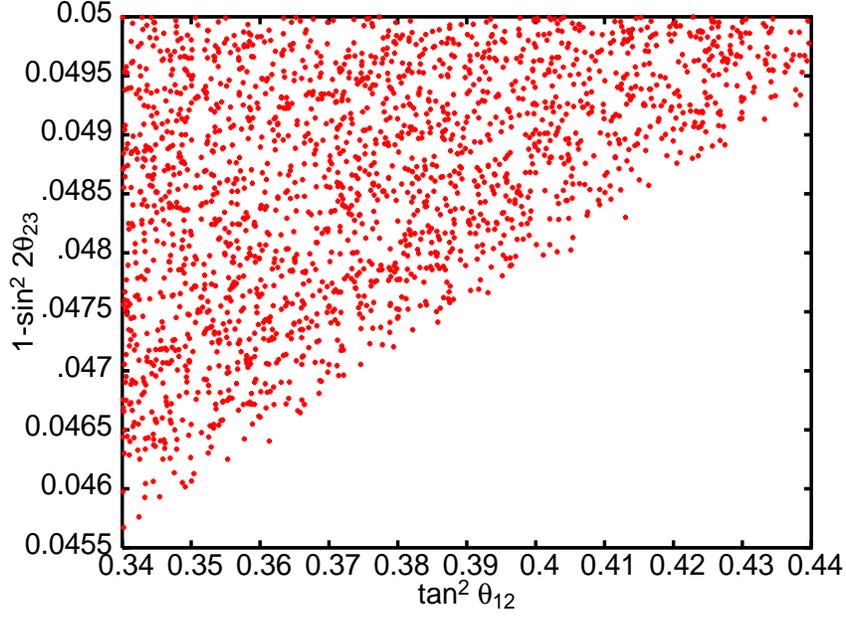,width=11cm,height=8cm}
\caption{\label{fig:c=0}Scatter plot of $\tan^2 \theta_{12}$ 
against $1 - \sin^2 2 \theta_{23}$ for the case of $\zeta_\nu=0$.} 
\end{center}
\end{figure}

\begin{figure}
\begin{center}
\epsfig{file=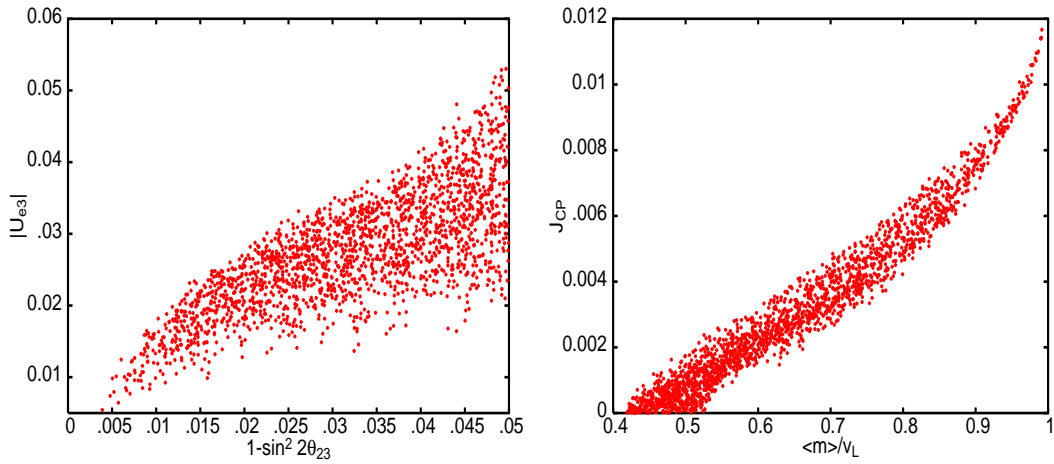,width=16cm,height=21cm}
\vspace{-13.5cm}
\caption{\label{fig:c=1}Scatter plot of $1 - \sin^2 2 \theta_{23}$  
against $|U_{e3}|$ and $\langle m \rangle/v^{~}_{\rm L}$ against 
$J_{\rm CP}$ for the case of $\zeta_\nu=1$.} 
\end{center}
\end{figure}

We first vary $\zeta_\nu$ between 0.2 and 10 since we expect from 
the above discussion that in this range $|r_\nu|$ will be of
${\cal O}(0.1)$ to ${\cal{O}}(1)$. Larger values of 
$\zeta_\nu$ will result in unnaturally large values of 
$c^{~}_{\rm D}$ as long as $\gamma$ is of order one. 
For the sake of simplicity, here we only take account 
of $r_\nu \ge 0$ but emphasize
that a similar analysis for the $r_\nu \leq 0$ case is straightforward. 
Furthermore, all small perturbative parameters 
appearing in $M_l$ and $M_\nu$ are allowed to vary between 0 and 0.2.
The relevant neutrino oscillation parameters are required to lie in the 
following ranges, which are the typical $1\sigma$ outcome of recent 
global analyzes \cite{bahcall,valle,STP}:  
\begin{eqnarray}
\tan^2 \theta_{12} & = & 0.34 \ldots 0.44 \; ,
\nonumber \\
|U_{e3}|^2 & \le & 0.015 \; ,
\nonumber \\
\sin^2 2 \theta_{23} & \ge & 0.95 \; ,
\nonumber \\
R_\nu & \equiv & \frac{\Delta m^2_{\odot}}{\Delta m^2_{\rm A}}
= 0.033 \ldots 0.053 \; .
\end{eqnarray}
We plot in Fig.\ \ref{fig:crbel} some of the resulting correlations 
between the model parameters and observables. 
It is seen that $r_\nu$ indeed is of ${\cal O}(1)$ for values 
of $\zeta_\nu$ larger than one. The functional 
behavior is excellently described by Eq.\ (\ref{eq:zetar}),
implying that the flavor democracy contribution to $M_\nu$ is
strongly suppressed due to significant cancellation between 
the conventional seesaw and triplet mass terms.
Regardless of the values of $r_\nu$ and $\zeta_\nu$, 
the neutrino mass ordering is of normal type. 
Moreover, the rephasing invariant of CP or T violation
$J_{\rm CP} = {\rm Im}\{U_{e1} U_{\mu 2} U_{e 2}^\ast U_{\mu 1}^\ast\}$ 
is positive\footnote{We also find a very fine-tuned region in 
the parameter space of $(\delta_{\rm D}, \delta_{\rm M},
\varepsilon^{~}_{\rm D}, \varepsilon^{~}_{\rm M})$, in which 
$|U_{e3}| \approx 0.1$ and $J_{\rm CP} \le 0$ hold. This possibility
seems quite unlikely and can be disregarded.}
and smaller than $\approx 1.2 \%$. This quantity measures the strength of
leptonic CP and T violation in neutrino oscillations. In addition,
the effective mass of the neutrinoless double beta decay
$\meff = \sum (m_i U_{ei}^2)$ is found to be 
of order of the common neutrino mass scale $v^{~}_{\rm L}$,
which may be at or below the level of ${\cal O}(0.1)$ eV. 
The deviation of $\sin^2 2\theta_{23}$ from one and that of 
$|U_{e3}|$ from zero are always non-vanishing. 
We see that the atmospheric neutrino mixing parameter 
$1 - \sin^2 2\theta_{23}$ is larger than $\approx 0.005$. 
On the other hand, $|U_{e3}|$ is also larger than $\approx 0.005$ but 
smaller than $\approx 0.06$. This upper limit is given by 
$|U_{e3}| \approx 2m_e/(\sqrt{6} ~m_\mu) \approx 0.057$, 
which is actually the prediction obtained from $\zeta_\nu = 0$ \cite{FX96}. 
For this special case, we show the correlation between 
$\tan^2 \theta_{12}$ and $1 - \sin^2 2\theta_{23}$
in Fig.\ \ref{fig:c=0}. It is clear that $1 - \sin^2 2\theta_{23}$
varies only slightly. Indeed, 
$\sin^2 2\theta_{23} \approx 8(1 + m_\mu/m_\tau + R_\nu\cos 2\theta_{12})/9 
\approx 0.95$ \cite{FX04}, which has nicely been reproduced by our
numerical analysis. For the case of $\zeta_\nu = 1$ we plot  
the correlations between $1 - \sin^2 2\theta_{23}$ 
and $|U_{e3}|$ as well as between 
$\langle m \rangle/v^{~}_{\rm L}$ and $J_{\rm CP}$
in Fig.\ \ref{fig:c=1}. The result for larger values of $\zeta_\nu$ is 
found to be essentially the same. Typically, larger values of $|U_{e3}|$ 
imply larger values of $1 - \sin^2 2\theta_{23}$ and less cancellation 
\cite{canc} in $\langle m\rangle$. On the other hand, 
$J_{\rm CP}$ becomes larger
when $\langle m\rangle$ approaches $v^{~}_{\rm L}$. 
Note that the numerical analysis only requires to reproduce the 
ratio of the solar and atmospheric neutrino mass-squared differences. 
Hence the common neutrino mass scale $v^{~}_{\rm L}$ is 
basically unspecified and ranges in our Ansatz from $\approx 0.06$ eV 
to $\approx 0.25$ eV, which is consistent with the limits from
laboratory experiments. Taking into account the most stringent 
cosmological limit on neutrino masses 
$m_i \leq 0.14$ eV \cite{WMAP} would cut the
afore-obtained upper bound of $v^{~}_{\rm L}$ by roughly a factor 
of two.

\vspace{0.2cm}

The question arises whether one can implement the scenario under study 
within a GUT framework. A typical problem will be that, e.g., the triplet 
term giving rise to $M_{\rm L}$ is associated with couplings that also 
contribute to the quark or charged lepton mass terms. Consider a 
renormalizable SO(10) theory with Higgs fields in the 10-plet and 
$\overline{126}$ representation. The relevant mass matrices in 
this case are given by \cite{so10}
\begin{eqnarray}
&& M_{\rm up} = v_{10}^{\rm up} Y_{10} + v_{126}^{\rm up} Y_{126} \; , ~~~~~~~~
M_{\rm down} = v_{10}^{\rm down} Y_{10} + v_{126}^{\rm down} Y_{126} \; ,
\nonumber \\
&& M_{\rm D} = v_{10}^{\rm up} Y_{10} -3 v_{126}^{\rm up} Y_{126} \; , ~~~~~~~
M_l = v_{10}^{\rm down} Y_{10} - 3 v_{126}^{\rm down} Y_{126} \; ,
\nonumber \\
&& M_{\rm L} = v^{~}_{\rm L} Y_{126} \; , ~~~~~~~~~~~~~~~~~~~~~~ \; 
M_{\rm R} = v^{~}_{\rm R} Y_{126} \; , ~~
\end{eqnarray}
with the Yukawa coupling matrices $Y_{10,126}$ and the vevs 
$v_{10,126}^{\rm up, down}$ for the up- and down-sector, respectively. 
To link this scenario with ours, $Y_{10}$ will have to correspond to the 
flavor democracy term. $Y_{126}$ will have to be this term plus a matrix 
proportional to the unit matrix. To assure that the latter term does not  
significantly contribute to the quark and charged lepton masses, the condition 
$v_{126}^{\rm up, down} \ll v_{10}^{\rm up, down}$ should be fulfilled. 
A detailed analysis of this situation is certainly interesting for the sake of
model building \cite{wz}, but it is beyond the scope of the present letter.

\vspace{0.2cm}

To summarize, we have combined the type-II seesaw mechanism with S(3) 
flavor symmetry and applied this idea to the neutrino phenomenology. 
Our starting point of view is that a Majorana neutrino mass 
matrix generally includes two terms allowed by S(3) symmetry, 
one being a purely democratic matrix and the other proportional to the 
unit matrix. As a consequence, for a conventional seesaw 
formula or a pure triplet term no large neutrino mixing 
can be generated.  For both cases the term proportional to the democratic 
matrix has to be highly suppressed. 
We have shown here that the suppression of this term 
can naturally be realized via cancellations in the type-II seesaw
scenario, from which the bilarge neutrino mixing pattern is in
turn achievable. For the explicit symmetry breaking Ansatz discussed
in this letter, we obtain a normal mass ordering, 
$0.005 \le |U_{e3}| \le 0.057$ and 
$1 - \sin^2 2\theta_{23} \ge 0.005$. Furthermore, we find 
$J_{\rm CP} \ge 0$ and $\langle m\rangle/v^{~}_{\rm L} \ge 40\%$.
These instructive results can be tested in a variety of forthcoming 
neutrino experiments.

\vspace{0.2cm}

One of us (Z.Z.X.) is grateful to W.L.\ Guo and J.W.\ Mei for helpful
discussions.
This work was supported in part by the EC network HPRN-CT-2000-00152
(W.R.) and by the National Natural Science Foundation of China (Z.Z.X.).

\newpage

\end{document}